# Triangular Dynamic Architecture for Distributed Computing in a LAN Environment

Mahmud Shahriar Hossain, Kazi Muhammad Najmul Hasan Khan,
M. Muztaba Fuad & Debzani Deb
Department of Computer Science & Engineering
Shahjalal University of Science and Technology, Sylhet-3114, Bangladesh.
{shahriar_9639, najmul_bd}@yahoo.com, {fuad, banya-cse}@sust.edu

***ABSTRACT***

*A computationally intensive large job, granulized to concurrent pieces and operating in a dynamic environment should reduce the total processing time. However, distributing jobs across a networked environment is a tedious and difficult task. Job distribution in a Local Area Network based on Triangular Dynamic Architecture (TDA) is a mechanism that establishes a dynamic environment for job distribution, load balancing and distributed processing with minimum interaction from the user. This paper introduces TDA and discusses its architecture and shows the benefits gained by utilizing such architecture in a distributed computing environment.*

## 1. INTRODUCTION

Triangular Dynamic Architecture (TDA) introduces a mechanism of distributed processing and parallel computation for balancing the workload among the idle machines of a network. The construction of TDA is accomplished by introducing an intelligent server that dynamically categorizes hosts and relates those hosts transparently in a local area network.

### 1.1 GOALS AND OBJECTIVES

In a distributed system, there might be thin clients [7], who possess least processing capability with a minimum resource allotment; in contrary, there might be high performance hosts with idle CPU time. All the machines will be properly balanced with equal workload when TDA is applied. A TDA server always haunts for job from all the clients. When the server finds a job from a client, it divides the job into granules and distributes it to the service providers. After processing, the service providers directly return the outcomes to the requesting client. An intelligent server must divide the requested jobs efficiently so that the distribution mechanism properly balances the load across the system. A fallacy may arise when the server is deciding which service providers are going to get the job. A service provider idle at this moment may not be idle a minute later. The server should always be ready to detect such issues to balance the load.

An access to a large database with billions of records is basically occurred through indices. Access points are mentioned in the index table for faster interpretation. Another consequence is that the index table in turn may grow larger. The databases in use may have different formats and may be dispersed at different geographical locations. A client requesting for a record is delayed if the search is linear. If indices are used, a portion of the database needs to be accessed linearly. Such a linear access becomes a cause of large access time. Undoubtedly, dividing the job into different machines reduces the access time. However the distribution of load itself can overwhelm the system. Hence a mechanism to resolve this problem is mostly necessary to implement the architecture. The architecture needs to be dynamic because a static architecture cannot always provide a perfect solution to the client requests. Moreover, platform independence is a vital issue for maintaining different formats of databases, as well as processing client requests arriving from different locations. The system should completely hide the entire mechanism of distribution from the user providing an easy interface.

Once again, computation intensive calculations like matrix multiplication can also be divided into pieces to different hosts so that the computation time is reduced. Communication overhead becomes critical in such a scientific calculation. Special delegation mechanism should be used to reduce congestion over the network. The critical aspect of the architecture is to provide different types of job distribution through the same server. The architecture should support platform independence because in a local area network hosts of different operating system and architecture exist. So, TDA is provided with the facility of Java Virtual Machine (JVM)[12]. Sophisticated distribution mechanism should be used for distributing the granulized job. TDA utilizes Remote Method Invocation (RMI)[13] for load distribution by reason of its flexibility of different types of object passing in an object-oriented fashion. Another critical aspect is that the hosts of a local area network varies in CPU speed, memory size, bus speed, background processes and many other parameters. TDA should homogenize (Section 3.5) the hosts to such a platform where the load is balanced in the network rather than providing only an equal distribution.

### 1.2 SCOPE

TDA enriches parallel and distributed processing mechanism in a sophisticated manner across a local area network. For large-scale distributed information systems where billions of records are processed, hundreds of users are served concurrently and large varieties of services are provided, TDA becomes a promising architecture for faster implementation. E-commerce based transaction processing managed by hundreds of hosts based on TDA will provide platform independence, faster transaction processing, ease of management and varieties of other services. TDA supports various types of business-oriented jobs as well as computation intensive scientific jobs to be distributed in a local area network.

### 1.3 RELATED WORKS

TDA is a major extension of the three-tier model [1]. Architectures like TDA should possess own mechanisms to resolve the task of distribution. Few other research projects [3, 4, 5, 6, 8, 9, 10, 11] have studied client-server, three-tier architecture and object distribution.

Scott [11] introduces the basics of client/server computing and component technologies and then proposes two frameworks for client/server computing using distributed objects. The component-based architecture defines the basic preliminary components of TDA. TDA is further developed to communicate among three kinds of hosts: server, client and service-provider. Moreover, TDA establishes dynamic relations on runtime.

Randall et el. [10] have discussed the scalability of a client server relationship. The distribution architecture is developed turn by turn as the number of clients is increased. The paper describes several existing distributed object oriented systems but they did not show any kind of performance measurement benchmarks against their comments.

JavaParty [9] transparently adds remote objects to Java [12] by declaration in the source code. It introduces involvement of pre-compiler. It creates multiple Java byte-code files for every single distributable class. TDA does not require any pre-compiler for its transparent distribution. JavaParty is specially targeted towards and implemented on clusters of workstations. It combines Java-like programming and the concepts of distributed-shared memory in heterogeneous networks. An aim of establishing TDA goes with a strict promise that no change of the existing compilers would be done. Neither any kind of hardware dedication would be used nor the network topology would be changed.

Another work experimentally compares mechanism of load balancing with existing load-balancing strategies that are believed to be efficient for multi-cluster systems. Nieuwpoort et al. [8] conducted this comparison and established a divide-and-conquer model for writing distributed supercomputing applications on hierarchical wide-area systems. But the mechanism in TDA is established in a local area network, without making any kind of change in the hardware level. Searching to a database should not be distributed in a divide and conquer strategy, because such a distribution makes the mechanism dependent on large varieties of hosts which would result in higher amount of communication overhead.

The aim of NODS [3] is to define an open and adaptable architecture that can be extended and customized on a per-application basis. Furthermore, services or database systems configuration can be adapted at runtime (e.g., add new services, change services, internal policies), according to environmental change. TDA also enables such runtime facilities and their improved versions depending on the TDA server.

Edelstein et al. [4] describes different client-server relationships whereas some other research works [1, 2] also argues different tier-concepts like three-tier and two-tier. A two-tier system disperses user system interface, some processing management and database-management, in two different layers. In three-tier model, the three-layers are user system interface, process management and database management. TDA is a further development of three-tier model that is capable of performing vast varieties of jobs in dynamic fashions other than database handling. Compared with static three-tier model, TDA is dynamic in nature.

Fuad et al. [5, 6] introduces a system called AdJava that harnesses the computing power of underutilized hosts across a LAN or WAN. It also provides load balancing and migration of distributed objects through the use of intelligent software agents. Although the migration mechanism used in AdJava is highly automated, it suffers from penalty of migration of objects. TDA provides mechanism to pass objects to the server and thereafter service providers, but there are administrative preferences that allow real distribution of load through analyzing it entirely or a virtual distribution of load that allows distribution information collection from the server. AdJava uses a simple distribution policy to distribute objects to available machines. If the number of objects to be distributed is more than the number of machines in the system, AdJava distributes more than one object to those machines that are loaded lightly compared to other machines in the system. But TDA distributes a computation according to the homogenized information about the system. Objects are granulized according to that dynamic information. So there is no need to recycle object-transfer to already loaded service providers by a granule of the same request. AdJava harnesses its performance only through scientific applications while TDA is capable of distributing business applications as well.

## 2. TDA OVERVIEW

TDA is a sophisticated form of client-server relationship that in turn is established over three-tier architecture. Now the classical client server relations are no more suitable[3], applications now follow the three-tier architecture. In TDA, the classical client-server relationship is established dynamically and the three-tier architecture is then merged to it. TDA offers a cyclic triangular relationship (CTR) which is dynamically established by the server. The relationship is constructed between the client, the server and the service provider. A service-providing host, for convenience, is called a Client Service Provider or CSP. As shown in Figure 1, a client makes request to the server to perform a job, the server then hands the job over to an appropriate CSP. After processing the request, the CSP directly sends the result to the requesting client.

### 2.1 TRIANGULAR DYNAMIC ARCHITECTURE (TDA)

TDA is called so because the CTRs are established on demand and dynamically at run time. For all of the triangles, the server serves as the common point. The server establishes CTRs when any request is made. The server may also decide to make several CTRs against a single request. Thus the efficiency of the TDA depends on how efficiently job requests are divided into granules and how the CTRs are established. The relationships can also switch from one to another, that is, if the CSP of a CTR becomes busy after receiving the sub-request from the server, it can send the server a connection refusal request and also sends the current status of the sub-job it was performing. If the server grants the refusal request then the CSP is free, the server will hand over the remaining part of the sub-job to another CSP that is least busy. It is evident from the Figure 1 that the server is the common point for all the triangles, which means that the server is the one who is responsible for establishing such relations. This is the basis of TDA.

If Client1 sends a request to the server and if the server decides that the request can be divided into three parts, it sends the granulized requests to three CSPs designated as CSP1, CSP2 and CSP3. The three CSPs process the corresponding sub-jobs in parallel and send the outcomes directly to Client1. In this case three CTRs are established, (i) client1, server, CSP1, (ii) client1, server, CSP2, (iii) client1, server, CSP3. For all these dynamically established CTRs, the server is the common element, which proves that server is the one that is responsible for the decision of distribution.

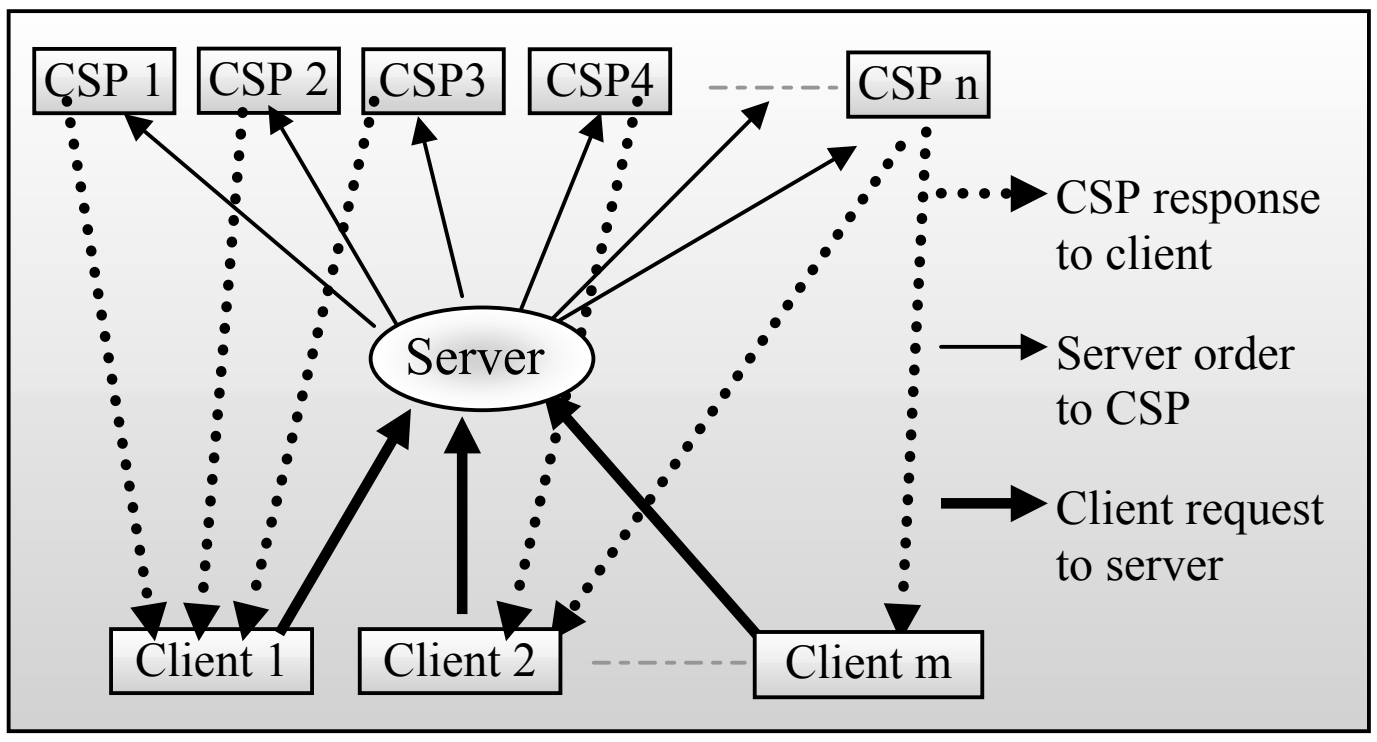

**Figure 1: Sample Triangular Dynamic Architecture.**

### 2.2 MERGING THREE-TIER TO TDA

If the three-tier model is rearranged such that, the middle tier of the three-tier model represents the CSPs in the TDA and a new layer is introduced that holds the TDA server (Figure 2), the combination gains several benefits. The TDA server gains control over the process management layer of the three-tier model. Each management section is now considered to be a CSP, hence the server gets the ability to distribute them jobs. This merging process favors subdividing a database query into sub-queries by the TDA server and therefore processing them in different CSPs for faster execution. Different CSPs can perform different jobs. The TDA server is informed about the abilities of the CSPs. For instance, some CSPs can read from a database, where others can upgrade it; there may several CSPs that are able to perform other computation intensive jobs. Some gaming ports may also exist where deep look ahead searching is involved like in chess. Figure 2 shows such a resulting architecture when TDA is merged with three-tier model.

The relation between CSP level to User System Interface level depends on the decided CTR established by the server. The relation is dynamic because the server on demand establishes CTRs in runtime. The CSP level has two parts – one corresponding to the middle tier of the three-tier model i.e. the middle tier CSPs and another corresponding to jobs other than database relation. The TDA server is the one to decide which portion is going to get the subdivided jobs. Therefore, such a merging can provide faster lookup to database, faster updates, with other facilities as provided by the CSPs.

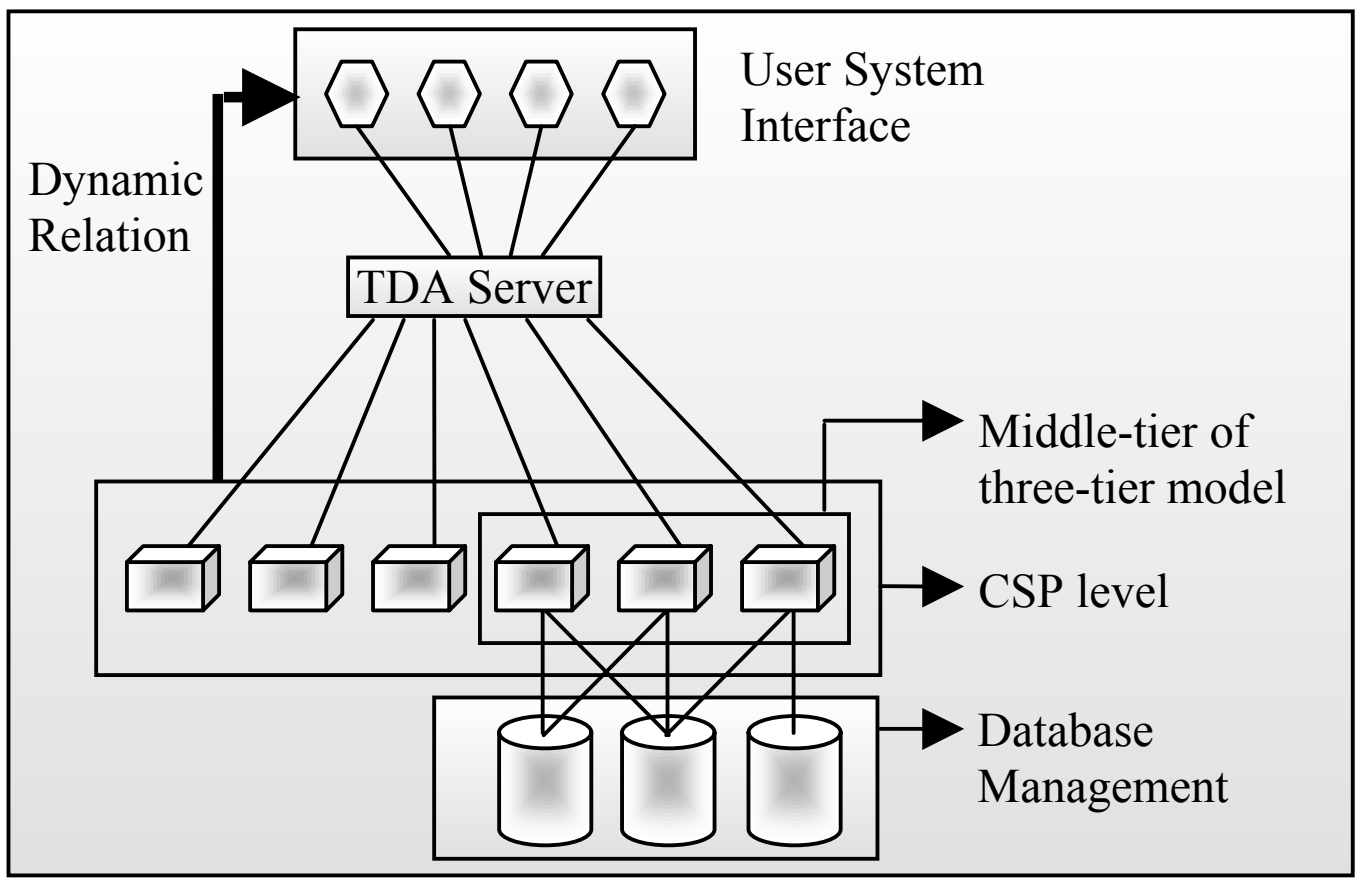

**Figure 2: Three-tier architecture lying inside TDA.**

## 3. IMPLEMENTATION DETAILS

Implementation of TDA is a challenging job. A vast variety of consequences are to be fulfilled while implementing TDA. This section describes the implementation details.

### 3.1 USING RMI FOR TDA

With Remote Method Invocation (RMI), Java objects that reside on different hosts can be used in a distributed fashion, i.e., remote object references provided by a name server are bound to local variables. Then methods can be called on those remote objects. RMI facilitates object function calls between Java Virtual Machines (JVMs). JVMs can be located on separate computers - yet one JVM can invoke methods belonging to an object stored in another JVM. Methods can even pass objects that a foreign virtual machine has never encountered before, allowing dynamic loading of new classes as required.

When a CSP is interested in connecting with the server it looks up for a registration query to the TDA server. The system is built up in such a manner that, the CSPs become interested during bootstrap; a tiny bootstrap loader is responsible for getting connected with the server. The loader just runs the program that communicates with the server to receive an ID for convenient handshaking mechanism. The loader performs its job when the machine is first turned on. Without a proper registration to the server, the server cannot send serialized objects holding job requests to the CSPs.

The discussion so far is about establishing a client-server relationship with the CSPs only. In the same manner, client-server relationship between client side programs and the server can be established using RMI. Implementing CTR using RMI is a challenging task. The challenge lies inside the reference passing mechanism, using RMI. The server passes a reference of the client side program to the CSPs along with the granule that the server has decided to hand over to a CSP. That is, whenever a client sends a request to the server it also sends its reference along with the request.

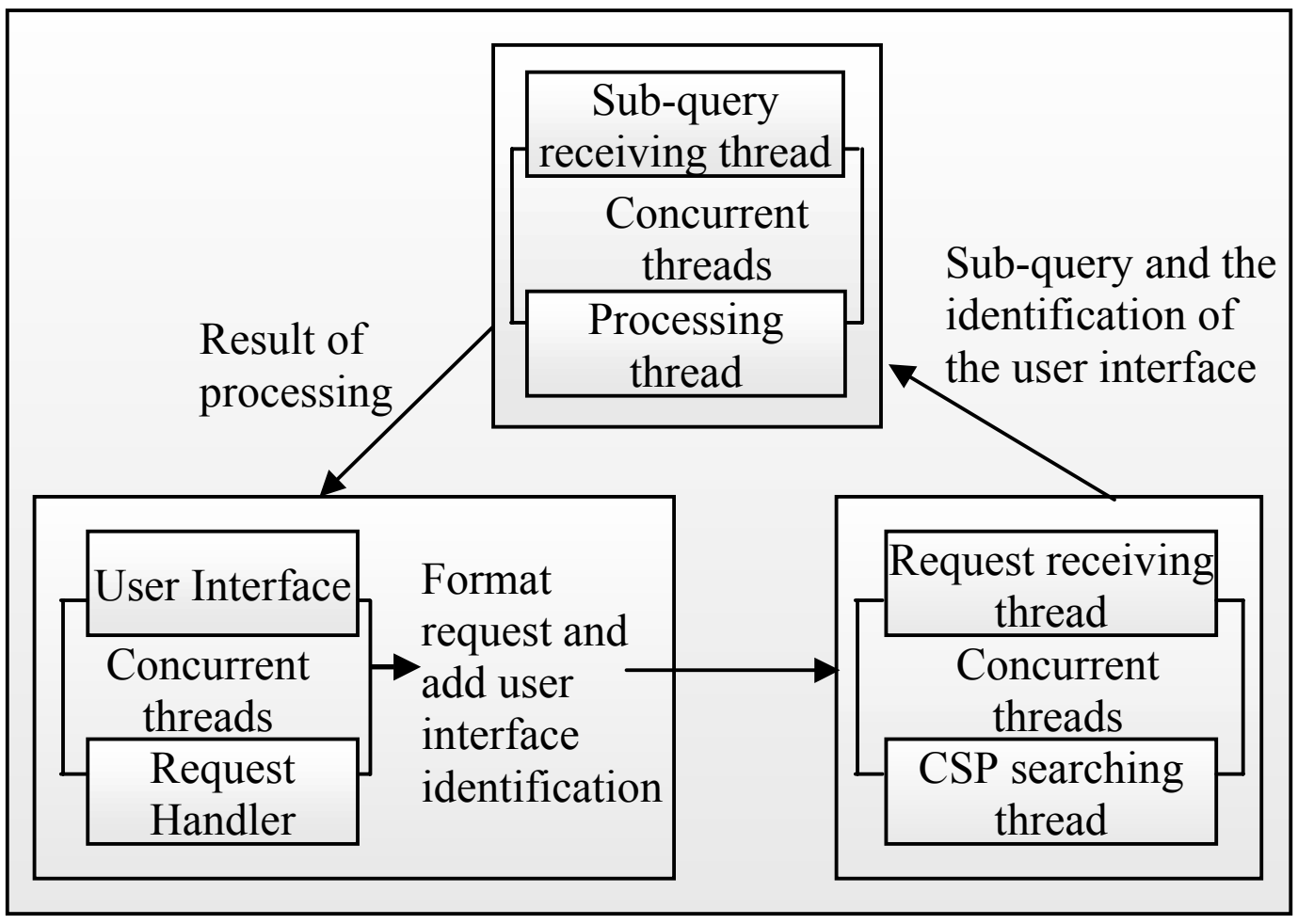

**Figure 3: Implementation of Cyclic Triangular Relationship (CTR) Using RMI.**

RMI passes parameters in two ways – pass by value and pass by reference of an object. A pass by value mechanism transfers the whole object to a remote host as parameter. But a reference parameter does not require the transfer of the whole object but only the access path to the object that is to be sent. This mechanism enables a CSP send the result after processing, directly to the client. The implementation using RMI is illustrated in Figure 3. The client side program has basically two parts - a user interface and a request handler. User requests are sent to the request handler of the client side program. The request handler encrypts user request and adds user program identification as a reference parameter.

The server receives the request along with the reference and finds out the most efficient division of the query. CSP searching thread in the server then sends sub-query to the CSPs. The CSPs grasp the sub-query, process corresponding sub-jobs and then directly send the result to the appropriate thin client - at specific reference. It should be mentioned that the programs for performing computation and database accesses inside CSPs are all background processes. All these are totally hidden from the user-knowledge of the particular CSP. Also, the requesting client does not have any idea about all the remote programs, threads, and other components.

### 3.2 TDA SERVER

TDA server is one, which is responsible for the actual distribution of workload. The server maintains some information and based on the stored information, the server can decide about the number of granules to be generated for a particular request. When a request arrives, the server always depends on the latest data available to its local database; it does not look for more information from the CSPs, since doing so will degrade its performance.

### 3.3 CSP

Background processes are the heart of CSPs. All the processes of a CSP are hidden from the remote user's sight. A background process always measures the current load of the computer even when the CSP is doing its share of the work. But, it measures its load in such a manner that it does not overwhelm other processes because it is implemented through a low priority thread. Time to time, it communicates with the server mentioning the current load. The server hence upgrades the local database about the corresponding CSP. If the remote host becomes busy with user task over a measured threshold, it tries to hand over the partially completed job. The server then searches for a new CSP that is least busy. If such a CSP is found, server permits the previous CSP a refusal after getting the status of the sub-task it was resolving. The rest of the sub-task is then performed by the new CSP, starting from the point where the previous CSP has left. If such a refusal is not possible then the CSP requesting a refusal must conduct the sub-task; in that case the server marks the CSP as a busy one until the background process form that CSP does not inform the server about its load reduction.

### 3.4 CLIENT

The overall TDA is designed to facilitate the client; to reduce the access time to a database and to perform many other jobs that the client alone was unable to conduct efficiently. Furthermore, the client might never perform the job as a thin host. A client program is composed of a user console and a request handler. User console is the basic interface to TDA for the users. If a user casts a request through the console, the request is sent to the request handler. Request handler encrypts the request and sends the request along with the client object reference to the TDA server. The result of processing is received in the user interface portion.

### 3.5 HOMOGENIZING TDA

The server maintains several tables in its local database that helps distributing the load. The server actually calculates the scope-length to be offered to a particular CSP, using the tables of the local database. Most critical knowledge-issues are performance of the CSP, their response time, list of services provided by a CSP, etc. A background process in the CSP informs the server about its current load after every 30 seconds. The server maintains this information and based on the stored information, the server generates a performance number, which is called the homogenized performance. The server depends on the homogenized performance of the CSPs for the balanced distribution of load.

Whenever a CSP gets an identity during bootstrap, it sends performance parameter to the server. The server also measures the communication distance of the CSP by pinging it test packets. Time to time, the server upgrades its tables e.g., it sends test packets to get the response time of the CSPs. Test packets are directly thrown back to the server from the CSP. Also it helps the server to know whether a particular CSP is dead or active. Test packets are smaller in size and they merely congest the traffic. If a CSP is not busy, but yet it has a large response time, then the server does not invoke it for small jobs. The server always tries to offer it massive and computation intensive jobs so that the time consumed by communication overhead becomes less pronounced.

A CSP with comparatively lower homogenized performance always gets smaller portions of request than a faster one. A service-provider that is dead with a sub-request keeping it incomplete is again re-requested to another CSP by the server. This prevents loss of sub-requests, hence the possible loss of client-request. Some CSPs are marked by the administrator as lazy and least busy all the times. Server prefers them as first priority to be involved by sub-requests. The administrator can also set a threshold value for homogenized performance. TDA server ignores CSPs that have homogenized performance less than the threshold value. Therefore, homogenization improves TDA to not only a distributing architecture but also a sophisticated load-balancing design.

## 4. PERFORMANCE ANALYSIS

TDA is able to perform distribution of different types of jobs. To verify the potentiality of TDA, a business application is introduced in this section. Performance is measured in three types of environment: heterogeneous environment, homogeneous environment and homogenized environment. A heterogeneous environment is one, where hosts of different configurations in hardware exist. They may also vary in their operating systems, background daemons, virtual memory allocations and many other parameters. A homogeneous environment consists of systems with same hardware configuration. It is hard to get a perfectly homogeneous environment because the hosts may have different background processes and applications. A homogenized environment is one where TDA has applied homogenization i.e. in reality, homogenized environment is a heterogeneous one, but TDA homogenized the overall system.

A linear search to a large database is a time consuming job. TDA offers searching from different locations of a table of records. A linear search-time can be reduced with concurrent search by multiple CSPs from different locations. The entire search area is divided by the TDA server and searches are conducted by CSPs from the server-decided locations.

### 4.1 SYSTEM APPARATUS

A global database with one hundred thousand records is constructed. The database is global to all the CSPs so that all can make access to it. A linear search by one host takes a significant amount of time to complete the search. This sample problem is solved by TDA to verify the effect of TDA on overall search time.

The experiments were taken with various combinations of Intel machines. They manifested a heterogeneous infrastructure. For experiment with homogeneous environment, six computers with same hardware configurations were selected. For heterogeneous and homogenized environment Pentium II, III, and IV Intel machines with physical memory ranging from 64 to 128 MB were used. All of them are connected by 100 Mbps Ethernet network. All the TDA components were running over the Virtual Machine provided by SUN's JDK version 1.2.2 or higher.

### 4.2 HOMOGENEOUS BEHAVIOR OF TDA

For better understanding, homogeneous behavior is illustrated first. Theoretically, application of TDA in a homogeneous environment should produce a linear speedup, but in reality an overhead is found which is a function of decision time made by the TDA server and communication distance. As a result the speedup would not increase linearly as newer CSPs are involved in a searching. Figure 4 shows the homogeneous behavior of TDA. Figure 4(a) shows the relative time required for overall search against different number of homogeneous machines involved. The overhead and actual searching time is also denoted in this figure. Corresponding speedup is plotted in Figure 4(b). It shows that without considering the overhead, speedup is almost a linear function (the gray line). The black line shows the actual speedup against different number of CSPs considering the overheads. This analysis shows that TDA reduces the actual computation time as newer CSPs are involved but the overhead flattened the expected linear behavior. For higher amount of load the overheads would become negligible and therefore the real speedup should become an almost linear function.

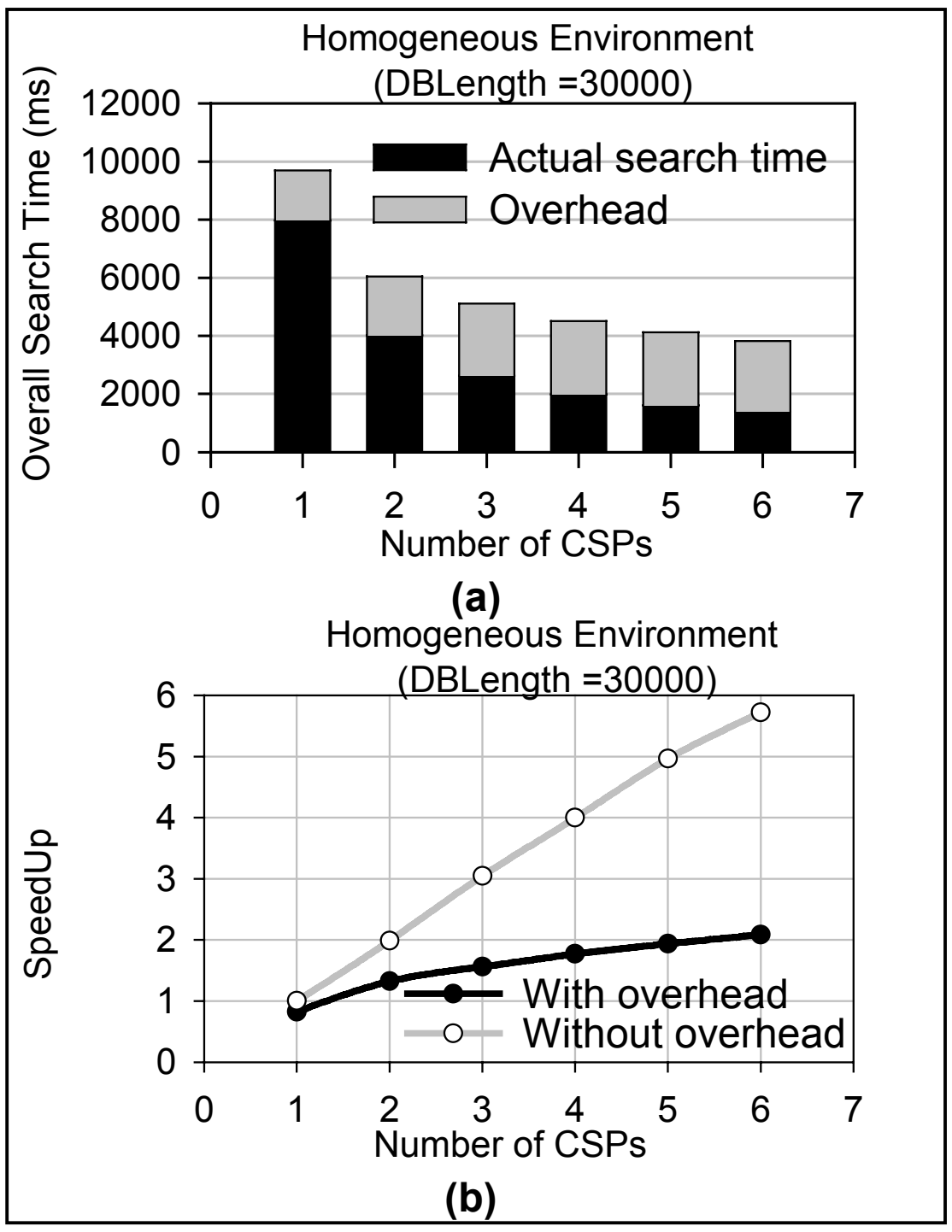

**Figure 4: Homogeneous behavior of TDA (linear search in 30000 records).**

### 4.3 HETEROGENEOUS BEHAVIOR OF TDA

If the TDA server does not have any homogenizing capability then a heterogeneous behavior is experienced. In a heterogeneous LAN environment, if the server distributes a load into equally granulized sub-jobs, an unpredictable performance is detected due to different CPU speed, operating system, amount of memory, etc. In reality, a network is configured with heterogeneous machines. So TDA should deal with heterogeneous environment with close observation.

Figure 5(a) shows the actual search time, overhead and thereafter the overall search time for using different numbers of CSPs. Subsequent addition of newer CSPs reduces the actual search time. But, addition of the sixth CSP and ninth CSP increases the actual search time because these two CSPs were comparatively of lower performance. Equal distribution of load to all the CSPs may suffer from this type of performance degradation. The corresponding speedup is shown in Figure 5 (c) with a gray line. Speedup is fallen when sixth and ninth CSPs are introduced to the distribution. Hence the speedup with equal distribution of load is mostly dependent on the low-performance machines, because the completion of the search is subject to the completion of the sub-request executed by the slowest CSP. Heterogeneous behavior of TDA introduces a massive problem of equal distribution for parallel computation. It expresses that there should be a mechanism that would enrich TDA with distribution of load according to the performances of the CSPs invoked for computation. All the CSPs should complete the partial computations almost at the same time. This is possible only when slower CSPs would get lower amount of load and the high performance CSPs will get larger portion of the job. This introduces the concept of homogenization.

### 4.4 HOMOGENIZED BEHAVIOR OF TDA

The same physical heterogeneous environment of Figure 5(a) is virtually homogenized by TDA and actual searching time is found to be always reducing. Figure 5(b) shows that actual searching time is always reducing although the sixth and ninth CSPs are slow and loaded. TDA server delivered lower amount of load to the slow and loaded machines. This introduces a sophisticated balancing mechanism of loads across a local area network. The corresponding speedup for the homogenized environment is plotted in Figure 5(c) with a black line. It shows that homogenization ensures speedup improvement upon addition of a newer CSP whereas equal distribution of load does not ensure performance improvement. Equal distribution suffers from heterogeneous behavior of parallel computation. Figure 5 (c) shows that maximum speedup for heterogeneous behavior is found to be 2.3 when the distribution is committed between five CSPs. Subsequent addition of newer CSPs improved the speedup for homogenized environment and maximum speedup was reached when nine CSPs were used. The maximum speedup gained during homogenized performance analysis was 2.8.

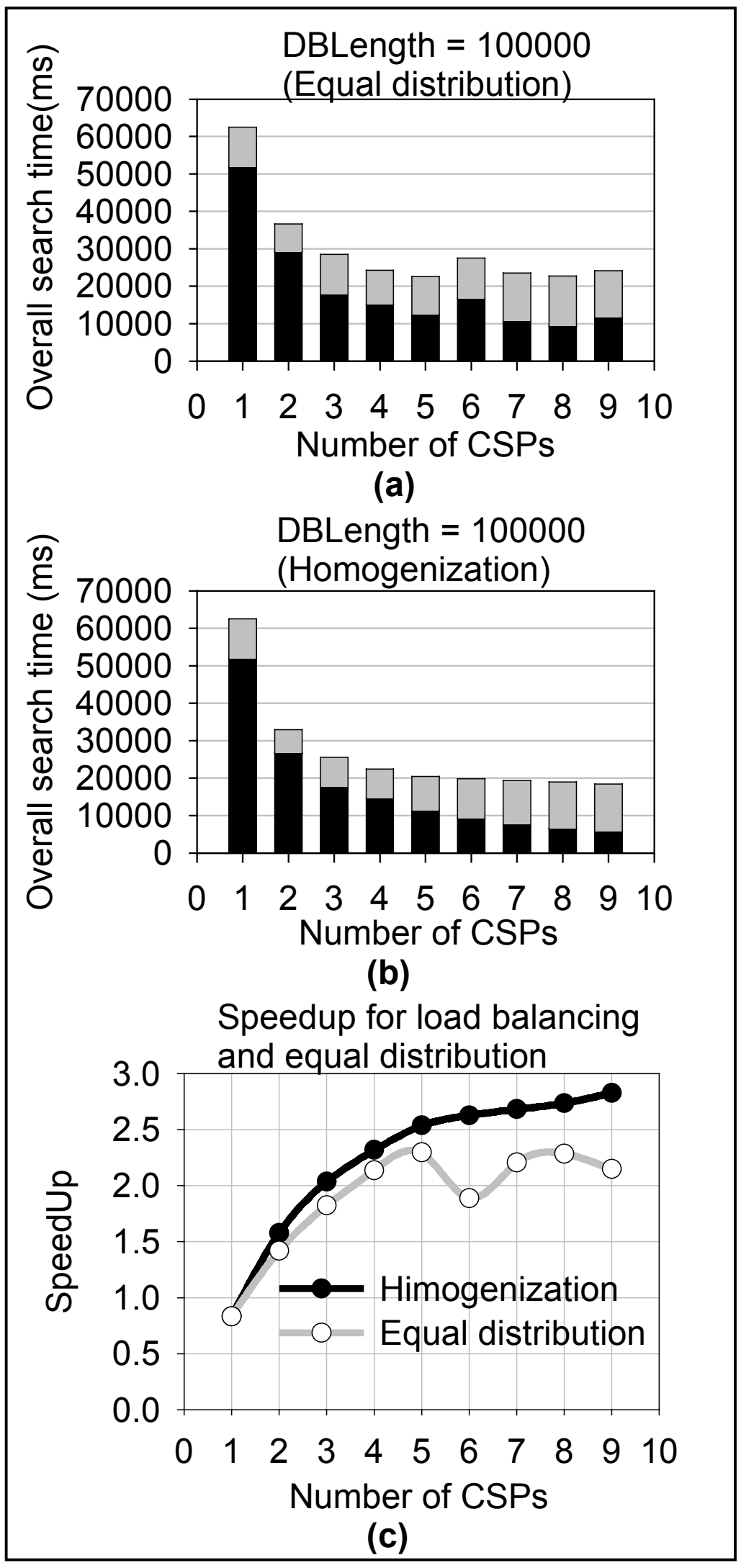

**Figure 5: Analysis with search across 100000 records (a) Heterogeneous behavior of TDA, (b) Homogenized behavior of TDA, (c) Corresponding speedup of (a) and (b).**

## 5. CONCLUSION

Triangular Dynamic Architecture is an enriching mechanism of job distribution across a local area network. TDA granulizes computation intensive jobs to concurrent pieces and operates them in a dynamic environment to reduce total processing time. TDA establishes dynamic load balancing and distributed processing mechanism with minimum interaction from the user.

TDA provides better processing time in a distributed computing environment. For implementing TDA, the present JVM remains unchanged. The current implementation is fully based on the existing JVM and that way TDA fulfills its main goal of providing a distributed computing environment in an existing LAN.

## 6. REFERENCES

[1] Carnegie Mellon Software Engineering Institute, “Three Tier Software Architectures”, http://www.sei.cmu.edu/str/descriptions/threetier.html.

[2] Carnegie Mellon Software Engineering Institute, “Two Tier Software Architectures”, http://www.sei.cmu.edu/str/descriptions/twotier.html.

[3] Collet Christine, “The NODS project, Networked Open Database Services”, *Proceedings of Symposium on Objects and Databases (ECOOP),* http://www-lsr.imag.fr/ Les.Personnes/Christine.Collet, LNCS 1813, Sophia Antipolis and Cannes, France, June 2000.

[4] Edelstein H., "Unraveling Client/Server Architecture", *DBMS* (http://www.dbmsmag.com), Vol. 7, No. 5, Page 34, May 1994.

[5] Fuad M. M. and Oudshoorn M. J., "AdJava - Automatic Distribution of Java Applications", *Australia Computer Science Communication*, Vol. 4, No. 28, Page 65-77, February 2002.

[6] Fuad M. M. and Oudshoorn M. J., “Automatic distribution and load balancing of Java objects in an agent oriented distributed system”, *ICCIT, Proceedings of 5th International Conference on Computer and Information Technology,* Page 101-107, Dhaka, Bangladesh, December 2002.

[7] Jennings T., “Application Deployment and Integration”, *Research Paper: Jacada ™ Ltd. Jacada ® for Java*, http://www.jacada.com/products/JacadabyButler.pdf, March 2001.

[8] Nieuwpoort R. V., Kielmann T. and Bal Henri E., “Efficient load balancing for wide-area divide-and-conquer applications”, *ACM SIGPLAN Notices*, Vol. 36, No. 7, Page 34-43, July 2001.

[9] Philippsen M. and Zenger M., "JavaParty – Transparent Remote Objects in Java", *Concurrency: Practice and Experience*. Vol. 9, No. 11, Page 1225-1242, November 1997.

[10] Randall L. Hyde and Fleisch Brett D., "A Case for Virtual Distributed Objects", *Int'l Journal on Parallel and Distributed Computing*, Vol 1, No. 3, September 1998.

[11] Scott M. Lewandowski, “Frameworks for component-based client/server computing”, *ACM Computing Surveys (CSUR)*, Vol.30, No.1, Page 3-27, March 1998.

[12] Sun Microsystems, “The Real-Time Specification for Java”, www.java.sun.com, 2003.

[13] Sun Microsystems, “Java Remote Method Invocation Specification”, www.java.sun.com/j2se/1.4.2/docs/guide/rmi/spec/rmiTOC.html, 2003.